\documentclass[10pt,english,preprint.aps,showkeys,showpacs]{revtex4}
\usepackage[T1]{fontenc}
\usepackage{float}
\usepackage{amsmath}
\usepackage{amssymb}
\usepackage{graphicx}
\usepackage{setspace}
\usepackage{esint}
\usepackage{color}

\makeatletter
\@ifundefined{textcolor}{}
{%
	\definecolor{BLACK}{gray}{0}
	\definecolor{WHITE}{gray}{1}
	\definecolor{RED}{rgb}{1,0,0}
	\definecolor{GREEN}{rgb}{0,1,0}
	\definecolor{BLUE}{rgb}{0,0,1}
	\definecolor{CYAN}{cmyk}{1,0,0,0}
	\definecolor{MAGENTA}{cmyk}{0,1,0,0}
	\definecolor{YELLOW}{cmyk}{0,0,1,0}
}

\@ifundefined{textcolor}{}{%
	\definecolor{BLACK}{gray}{0}
	\definecolor{WHITE}{gray}{1}
	\definecolor{RED}{rgb}{1,0,0}
	\definecolor{GREEN}{rgb}{0,1,0}
	\definecolor{BLUE}{rgb}{0,0,1}
	\definecolor{CYAN}{cmyk}{1,0,0,0}
	\definecolor{MAGENTA}{cmyk}{0,1,0,0}
	\definecolor{YELLOW}{cmyk}{0,0,1,0}
}

\usepackage{paralist}
\def\btt#1{\texttt{\@backslashchar#1}}
\DeclareRobustCommand{\bblash}{\btt{\@backslashchar}}
\usepackage{babel}
\usepackage{babel}
\@ifundefined{showcaptionsetup}{}{%
\PassOptionsToPackage{caption=false}{subfig}}
\usepackage{subfig}
\makeatother
\usepackage{babel}
\begin{document}

\title{Quantum Spin-5/2 Blume-Capel  Model in a  Random Transverse-Crystalline Field Anisotropy}

\author{Claudio M. Salgado}
\email{claudio_marcelo@fisica.ufmt.br}
\affiliation{Instituto de F\'isica, Universidade Federal de Mato Grosso, 78060-900, Cuiab\'a, Mato Grosso, Brazil.}

\author{Karollaine C. Leite }
\email{karolaine@fisica.ufmt.br}
\affiliation{Instituto de F\'isica, Universidade Federal de Mato Grosso, 78060-900, Cuiab\'a, Mato Grosso, Brazil.}

\author{Thiago M. Tunes}
\email{thiagotunes@fisica.ufmt.br}
\affiliation{Departamento de Ci\^encias Exatas e da Terra - Faculdade de Ci\^encia e Tecnologia, Universidade Federal do Mato Grosso, 78060-900, V\'arzea Grande, Mato Grosso, Brazil.}

\author{Marcelo F. Z.  de Arruda}
\email{marcelofza@fisica.ufmt.br}
\affiliation{Instituto de Federal de Mato Grosso, Campus de Sorriso, 78890-000, Sorriso, Mato Grosso, Brazil.}

\author{Jorge L. B. de Faria}
\email{hulk@fisica.ufmt.br}
\affiliation{Faculdade de Arquitetura, Engenharia e Tecnologia, Universidade Federal de Mato Grosso, 78060-900, Cuiab\'a, Mato Grosso, Brazil.}

\author{Alberto S. de Arruda}
\email{aarruda@fisica.ufmt.br}
\affiliation{Instituto de F\'isica, Universidade Federal de Mato Grosso, 78060-900, Cuiab\'a, Mato Grosso, Brazil.}

\begin{abstract}
In this work, we investigate the thermodynamic properties of the quantum Blume-Capel model with spin \( S = 5/2 \) in the presence of transverse and random crystalline fields. The system is described by a Hamiltonian that includes ferromagnetic exchange interactions between nearest neighbors, a longitudinal single-ion anisotropy, and a transverse single-ion anisotropy. Using a mean-field approach based on Bogoliubov's inequality for the Gibbs free energy, we derive the fundamental thermodynamic potential and the equation of state for the magnetization. The influence of the longitudinal and transverse anisotropy parameters on the magnetic ordering and phase transitions is analyzed in detail. We present magnetization versus temperature diagrams for various combinations of the anisotropies, exploring both positive and negative values. Our results reveal that the system exhibits standard second-order phase transitions for most parameter ranges, with no evidence of tricritical behavior. However, for certain positive values of the anisotropies, the model displays a first-order phase transition within the ordered phase, characterized by a jump from a higher-spin ordered state to a lower-spin ordered state. The critical temperatures are shown to be sensitive to the magnitude and sign of the anisotropy parameters. In particular, negative transverse anisotropies favor magnetic order, raising the critical temperature, while positive anisotropies promote disorder, lowering the critical temperature. This study provides a comprehensive analysis of the phase diagram of the \( S = 5/2 \) quantum Blume-Capel model and highlights the role of transverse fields in modifying the critical behavior.
\end{abstract}

\keywords{Blume-Capel model; transverse crystalline field; mean-field theory; phase transitions; magnetization; quantum spin systems.}

\pacs{ PACS:
05.30.-d, 
05.70.-a , 
05.70.Ce, 
05.70.Fh. }

\maketitle
\section{Introduction}

The study of magnetic systems with spin degrees of freedom greater than \( S = 1/2 \) has been a subject of continuous interest in condensed matter physics, primarily due to their rich phase diagrams and the complex interplay between exchange interactions, temperature, and single-ion anisotropies. Among these models, the Blume-Capel (BC) model stands out as a paradigmatic framework for investigating the effects of crystal fields on the thermodynamic behavior of magnetic systems. Originally proposed by Blume and Capel\cite{blu,cap}, this model incorporates a single-ion anisotropy term that favors or disfavors certain spin states, leading to the possibility of first- and second-order phase transitions, as well as tricritical points. These characteristics are not merely theoretical motivations, but have already been observed experimentally in compounds such as $FeCl_2$, $NiCl_2$, and in certain rare-earth-based magnets, where the crystal field splitting plays a dominant role in defining the magnetic response \cite{exp0,exp1,exp2,exp3,exp4,exp5,exp6,exp7,exp8,exp9}.

In addition to experimental results, the BC model also served as a testing platform for conflicting theoretical questions, such as the role of disorder in low-dimensional systems. The introduction of random crystal fields (longitudinal or transverse)\cite{clau1,seto1,ita0} suppresses long-range order in certain regimes; conversely, in other regimes it can stabilize exotic phases such as the Bose glass phase in magnetic analogues of disordered superfluids. Random transverse fields, specifically, break usual symmetries in a non-trivial way, often requiring advanced techniques beyond standard mean-field approaches, such as the replica method or quantum Monte Carlo simulations\cite{seto2}.

In its classical formulation, the BC model has been extensively studied using a variety of analytical and numerical techniques, revealing a rich critical behavior\cite{181,171,172,173,174,a2,cdd,a1,ku,k1,a3,kenz,ita1,ze1,ze2,mel,gul}. However, the inclusion of quantum effects, particularly through transverse crystal fields, introduces new features. The quantum version of the Blume-Capel model, which considers non-commuting spin operators, has received increasing attention due to its relevance in describing real materials, such as compounds with strong spin-orbit coupling or those subjected to external transverse fields.

Most studies involving quantum BC models have focused on lower spin values, particularly \( S = 1 \)\,\cite{ita3,so1,pla3,san1,so2}, due to the complexity associated with higher-dimensional Hilbert spaces. Nevertheless, systems with higher spins, such as \( S = 5/2 \), are of great importance as they can more accurately represent a wide class of magnetic materials, including those with high-spin transition metal ions. In these systems, the competition between longitudinal and transverse anisotropies can lead to novel ordering phenomena, including reentrant behavior and multiple phase transitions.

In this work, we investigate the thermodynamic properties of the quantum Blume-Capel model for spin \( S = 5/2 \) in the presence of both longitudinal and random transverse crystalline fields. We employ a mean-field approach based on Bogoliubov’s inequality for the Gibbs free energy, which provides a variational treatment of the free energy functional and allows us to derive the magnetization equation of state. The model incorporates two distinct single-ion anisotropies: a longitudinal term \( D (S_i^z)^2 \) and a transverse term \( D^x (S_i^x)^2 \). Our primary goal is to elucidate how the competition between these anisotropies and temperature affects the magnetic ordering, focusing on the nature of the phase transitions (first- or second-order) and the presence or absence of tricritical behavior.

The remainder of this paper is organized as follows. In Sec. II, we present the model Hamiltonian and describe the mean-field formalism used to compute the free energy and magnetization. In Sec. III, we present and discuss the numerical results, including a detailed analysis of the magnetization versus temperature diagrams for various combinations of the anisotropy parameters. Finally, Sec. IV is devoted to our conclusions and final remarks.

\section{The model and methodology}
\hspace{0.6cm}
This study aims to investigate the effects of transverse and random crystalline fields on the thermodynamic behavior of the Blume-Capel Quantum Model of Spin S = 5/2. Therefore, the system is described by the following Hamiltonian:
\begin{eqnarray}\label{eq.1}
{\mathcal H} = - J\sum\limits_{\left\langle {i,j} \right\rangle} {S_i^z}	S_j^z  -  D\sum\limits_i^N {{{\left( {S_i^z} \right)}^2}}  -	{D^x}\sum\limits_i^N {{{\left( {S_i^x} \right)}^2}}.
\label{h1}
\end{eqnarray}
Here, $S_z$ and $S_x$ are Pauli spin operators located at all sites in the lattice, and N is the total number of sites. $J > 0$ is the parameter that ferromagnetically couples all pairs of spins first neighboring. $D$ is the crystalline field on the z-quantization axis, and $D^x$ is the transverse crystalline field on the x-axis. The first term represents the magnetic interaction between the spin's first neighboring, the first sum is performed over all pairs $(zN/2)$ of the spins' first neighbors, and $z$ is the lattice coordination number.  The second and third terms represent the interactions between the spins and the crystalline fields at all lattice points. Thus, the second and third sums are performed on all spins located at all lattice points.   
\par
In Hamiltonian (\ref{h1}), one can directly observe some particular cases.  The first case, $D^x=0$, we recover the classical Blume-Capel (BC) model, therefore $D^x \not=0$ transforms the classical model of BC into the quantum model of BC. Second case, $D = D^x=0$, we gain the pure spin-5/2 Ising model, whose critical temperature, in the mean-field approach, is given by $T_c = 2/3$.
\par

In order to facilitate comparison with experimental results, we have used the spin identity $S(S+1)=(S_i^x)^2+(S_i^y)^2+(S_i^z)^2$ in Hamiltonian (\ref{h1}), from which we obtain:
\begin{eqnarray}\label{h2}
{\mathcal H} = - J\sum\limits_{\left\langle {i,j} \right\rangle } {S_i^z}
S_j^z - {D^x} \sum\limits_i^N {{{\left( {S_i^x} \right)}^2}}  -
{D^y}\sum\limits_i^N {{{\left( {S_i^y} \right)}^2}}, 
\end{eqnarray}
where irrelevant constants have been neglected.
\par
To investigate the effects of transverse crystalline fields on the magnetic properties of the quantum Blume-Capel model (BCQ) of spin-$5/2$, the mean field approximation via the variational method of Bogoliubov's free energy inequality will be used\cite{falk}. This approach has been widely used in several papers \cite{clau1,ita1}, and so the Gibbs free energy is given by the following inequality:
\begin{eqnarray} 
{\mathcal 
G}\le{\mathcal G_0} + \left\langle {\mathcal H} - {\mathcal H_0} \right\rangle_0 \equiv
\Phi \left( \eta  \right). \label{g1}
\end{eqnarray} 
In this context, ${\mathcal H}$ denotes the Hamiltonian (\ref{h1}) governing the system under investigation. ${\mathcal H_{0}}$ is a provisional Hamiltonian, representing a system of spins not interacting, with a known exact solution, and it introduces a variational parameter $\eta$. ${\mathcal G}$ stands for the free energy of the model described by the Hamiltonian ${\mathcal H}$, while ${\mathcal G_{0}}$ represents the free energy associated with ${\mathcal H_{0}}$. The brackets $\langle \cdots \rangle$ denote the thermal average over the canonical ensemble defined by ${\mathcal H_{0}}$. The minimum of $\Phi \left( \eta \right)$ concerning $\eta$ provides the approximate free energy, i.e., ${\mathcal G}= \Phi \left( \eta \right)_ { min}$.
\par
The choice of the tentative Hamiltonian is essential for the accuracy of the results obtained and has been extensively discussed in the paper\cite{ita1}. Then, the basic idea of a tentative hamiltonian is that it must
be as close as possible to the real Hamiltonian, the one that
wants to study. So, if ${\mathcal H_{0}} $ equals ${\mathcal H}$, it implies that
${\mathcal G}= {\mathcal G_{0}}$, and we have an exact result, which cannot be
obtained. So, we consider a trial hamiltonian ${\mathcal H_{0}} $ with $N$ spins, which only interact with an average effective  field. The simplest choice is given by:
\begin{eqnarray}\label{h0}
{\mathcal H_0}\left( \eta  \right) = - \eta \sum\limits_i {\left( {S_i^z}\right) - } \sum\limits_i {{D_i^x}{{\left( {S_i^x} \right)}^2}}  -\sum\limits_i {{D_i^y}{{\left( {S_i^y} \right)}^2}}.
\end{eqnarray}
Following the procedure explicitly described in equation (\ref{g1}), the variational free energy is expressed as:
\begin{eqnarray}\label{eq.14}
\Phi\left(\eta\right) &\equiv& {\mathcal G_0}+ \left\langle-J\sum\limits_{\langle {i,j} \rangle } {S_i^z} S_j^z- \sum\limits_i D_i^x\left( S_i^x \right)^2  - \sum\limits_i D_i^y\left(S_i^y \right)^2 \right\rangle_0\nonumber   \\
&+&  \left\langle\eta\sum\limits_i S_i^z  + \sum\limits_i D_i^x\left(S_i^x\right)^2  + \sum\limits_i D_i^y\left( S_i^y \right)^2 \right\rangle_0, 
\end{eqnarray}
or simply
\begin{eqnarray}\label{eq.30}
\Phi \left( \eta  \right) &\equiv& {\mathcal G_0}  - J zN\frac{m^2}{2}  
\end{eqnarray}
where ${\mathcal G_0}$ is defined by ${\mathcal G_0}=- K_BT\ln{Z_o(\eta) }$ and $m=\langle S_i^z\rangle $ or $m^2= \langle S_i^z S_j^z\rangle =  \langle S_i^z\rangle \langle S_i^z\rangle $, since in the mean-field idea, correlations between spins are neglected (spins are statistically independent). Therefore, the calculation of $Z_0$ is essential and is given by: 
\begin{eqnarray}\label{eq.38}
	Z_o &=& Tr\left\{ e^{\beta A(\nu)} \right\},
	\label{7}
			\end{eqnarray}
where  $A(\nu) = - {\mathcal H_{0}} $. Now, Introducing the relations of quantum mechanics:			
\begin{eqnarray}\label{eq.39}
S_i^+ &=& S_i^x + i S_i^y,\,\,\,\,\,;\,\,\,\,\, S_i^- = S_i^x - i S_i^y, \nonumber\\
 i\hbar S_i^z &=&  [S_i^x,  S_i^y] \,\,\,\,\,;\,\,\,\,\, [S_i^x,  S_i^z] = i\hbar S_i^y      \,\,\,\,\,;\,\,\,\,\, [S_i^y,  S_i^z] = i\hbar S_i^x\nonumber\\
S^{+}\mid s, m > &=& \hbar\sqrt{(s-m)(s+m+1)}\mid s, m+1>\nonumber\\
S^{-}\mid s, m >  &=&  \hbar\sqrt{(s+m)(s-m+1)}\mid s, m-1>,
\end{eqnarray}
through which the matrix $A(\nu)$ is determined:
\begin{eqnarray}\label{eq.118}
A(\nu)&=&\frac{1}{4}{\begin{bmatrix}10\eta+ 5\delta&{0}&{2\sqrt{10}\Delta}&{0}&{0}&{0}\\{0}&6\eta+ 13\delta&{0}&6\sqrt{2}\Delta&{0}&{0}\\2\sqrt{10}\Delta&{0}&2\eta+ 17\delta&{0}&6\sqrt{2}\Delta&{0}\\{0}&6\sqrt{2}\Delta&{0}&-2\eta+ 17\delta&{0}&2\sqrt{10}\Delta\\{0}&{0}&6\sqrt{2}\Delta&{0}&-6\eta+ 13\delta&{0}\\{0}&{0}&{0}&2\sqrt{10}(\Delta&{0}&{{-10\eta+ 5\delta}}\end{bmatrix}},\nonumber 
\end{eqnarray}
where $\delta= (D_x+D_y)$ and $\Delta= (D_x-D_y)$.
The diagonalization of $A(\nu)$, $det(A - \lambda I) =0$, provides the eigenvalues ($ \lambda_1$, $\lambda_2 $ , $ \lambda_3 $,  $ \lambda_4 $, $ \lambda_5 $, $ \lambda_6 $ ) that make it possible to obtain $Z_0$.
\begin{eqnarray}\label{eq.143}
Z_0= \sum\limits_{i=1}^6 {e^{\beta\lambda_i}}.
\end{eqnarray}
From $Z_o$, obtain the free energy associated with ${\mathcal H_{0}} $, which is given by:
\begin{eqnarray}\label{eq.150}
	{\mathcal G_0} = - NK_BT\ln{Z_o}.
\end{eqnarray}

With the free energy ${\mathcal G_0}$, we obtain the free energy given by equation  (\ref{g1}):
\begin{eqnarray} 
{\mathcal 
G} = {\mathcal G_0} + \left\langle {\mathcal H} - {\mathcal H_0} \right\rangle_0,
 \label{g11}
\end{eqnarray} 
where the variational parameter was determined by minimizing the free energy, $\eta = Jzm$. We also introduced other parameters: 
$\tau = \frac{1}{\beta J z}$, $\delta_{x}= \frac{D_x}{Jz}$, $\delta_{y}= \frac{D_y}{Jz}$.
The magnetization is obtained by minimizing the free energy with respect to the variable $m$, that is, $\frac{\partial {\mathcal G}(m)}{\partial m} = 0$. Due to the excessively large size of the expressions for the eigenvalues ($\lambda_i$), the partition function ($Z_0$), the free energies (${\mathcal G_0}$ and ${\mathcal G}$), and the magnetization ($m$), we do not write them explicitly.
\section{Results: }
In this section, we will elucidate the impact of transverse crystalline fields on the thermodynamic behavior of the Blume-Capel Quantum Spin Model with $S=5/2$. To undertake this analysis, we will present several diagrams depicting magnetization versus temperature ($\tau, m$) for various values of $\delta_x$ and $\delta_y$. Within our model, four Hamiltonian parameters ($\tau, J, \delta_x, \delta_y$) compete to govern the system's thermodynamic behavior. The temperature consistently induces disorder in the spin system through thermal agitation. Conversely, the positive coupling parameter (J > 0) between adjacent spins always tries to align the system ferromagnetically.
Simultaneously, the effects of single-ion anisotropies $\delta_x$ and $\delta_y$ come into play. When positive, these anisotropies favor spins assuming larger states (5/2) along their respective quantization axes, and when negative, they prefer smaller states (1/2). Consequently,   it contributes to both ordering and disordering the system.
\subsection{Pure Case: Spin-5/2 Ising Model }
Considering the generic nature of all our equations, we will initially focus on a particular case where the anisotropies are set to zero $(\delta_x = \delta_y=0)$  in the equation (\ref{g11}). This corresponds to the Spin-5/2 Ising model (pure case):
\begin{eqnarray}\label{eq.167}
 {\mathcal G}(m)&=&\frac{1}{2\tau}{m^{2}}+\ln  \left( 2\,\cosh \left( \frac{1}{2}\,\frac{m}{\tau}	\right) +2\,\cosh \left( \frac{3}{2}\,\frac{m}{\tau} \right) +2\,\cosh \left( \frac{5}{2}\,\frac{m}{\tau} \right)  \right) ,
\end{eqnarray}
 and $\frac{\partial {\mathcal G}(m)}{\partial m} = 0$, we have a magnetization is given by:
\begin{eqnarray}\label{eq.160}
m &=& \frac{1}{2}\left[ \frac{5\,\sinh \left( {\frac {5m}{2\tau}} \right) + 3\sinh \left({\frac {3m}{2\tau}} \right) +\sinh \left( {\frac {m}{2\tau}}\right) }{  \cosh \left( {\frac {5m}{2\tau}} \right) +\cosh \left( {\frac {3m}{2\tau}} \right) +\cosh \left( {\frac {m}{2\tau}} \right)  }  \right].
\end{eqnarray}
Curve 1 of Fig. 1(a,c),  and curves 8 and 3 in Fig. 1(b,d), depicted by dashed lines (red), illustrates the $\delta_x = \delta_y = 0$ representing the Pure Case: Spin-5/2 Ising Model. This curve presents the standard magnetization behavior, which indicates a second-order phase transition, ending at a critical temperature ( $\tau_c= 2.917$). By standard magnetization behavior (SMB), it is understood that the magnetization is maximum at zero temperature ($\tau = 0$), that is, in the absence of thermal agitation, the coupling interaction ($J > 0$) between the neighboring first spins is sufficient to maintain the spin system aligned in parallel (ferromagnetically). However, when the temperature increases, thermal agitation inverts the orientation of some spins so that the magnetization leaves its maximum value and goes continuously to zero at a critical temperature, where the phase transition (fe-pa) occurs. This SMB characterizes  a phase transition of second-order.  This curve  will be shown in subsequent diagrams as a guide for the eye.
\begin{figure}[htb]
	\centering
	\includegraphics[scale=0.5, angle=-00]{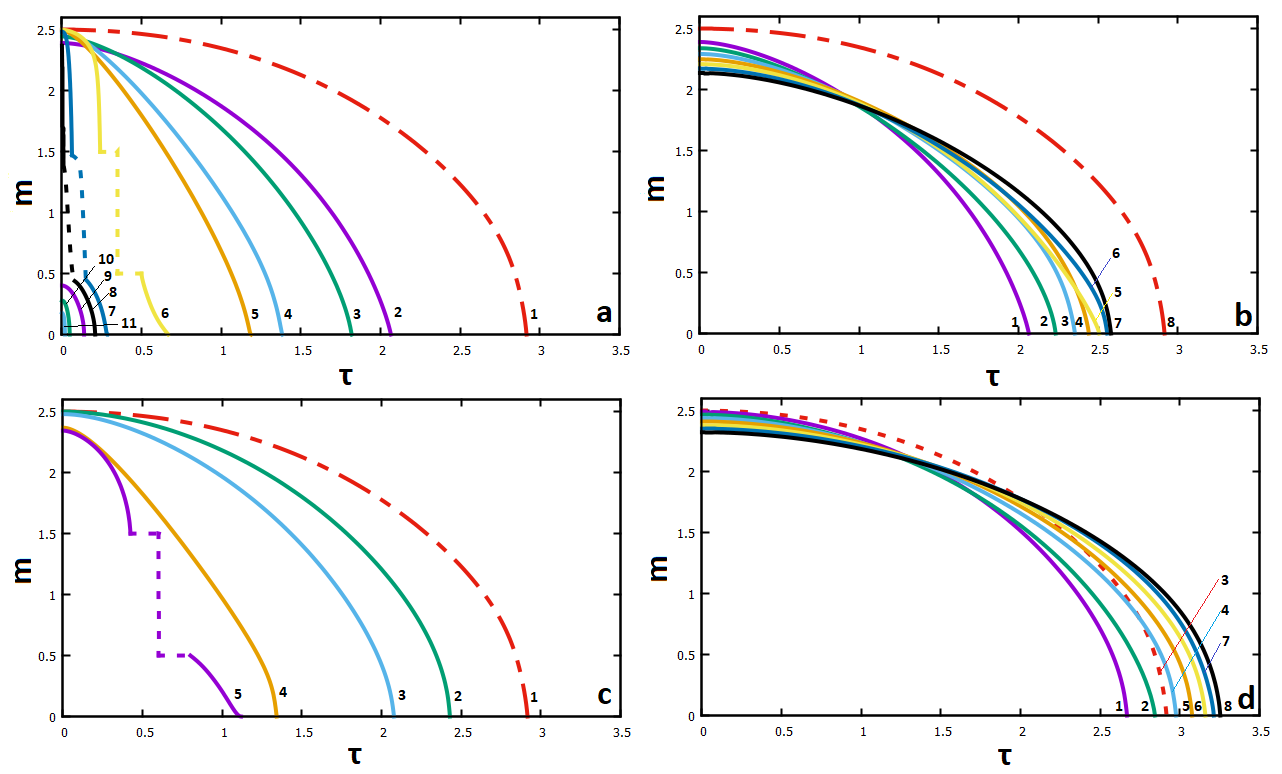}
	\caption{\it Diagram of magnetization versus temperature ($m - \tau$ plane). Figure (a) presents the effects of single-ion anisotropies with $\delta_y = 0.5$ fixed and $\delta_x$ varying. Curve 1 $\delta_y = 0.5 , \delta_x=0.0$ - dashed and dotted  line - pure model. The curves from 2 to 11 $\delta_x=0.0$ the anisotropies have $\delta_y = 0.5$ and $\delta_x = 0.0, 0.2, 0.4, 0.45, 0.5, 0.6, 0.7, 0.8, 1.0, 1.2$, respectively.  Figure (b) corresponds to $\delta y=0.5$ and negative $\delta$, varying $\delta_x = -1.2, -1.0, -0.8, 0.6, - 0.4, - 0.2, 0.0$ referring to curves 1 to 7, respectively. Figure (c) corresponds to $\delta_y=0.2$ and $\delta_x =
0.0, 0.2, 0.4, 0.6, 0.62$ linked to curves 1 to 5, respectively. Figure (d) corresponds to $\delta_y=0.2$ and $\delta_x =
0.0, -0.2, -0.4, -0.6, -0.8, -1.0, -1.2$ linked to curves 1;2;4;5;6;7;8 respectively.
	 Solid lines indicate standard magnetization behavior (SMB) and are related to a second-order phase transition. The dashed lines represent first-order phase transition within the ordered phase.}
	\label{f1}
\end{figure}
\subsection{Magnetization versus Temperature: Case $\delta_y=0.5$ Fixed}
 {\bf  Case $\delta_x$ positive:} 
Figure (1a) shows the magnetization behavior about temperature, with a fixed value of $\delta_y=0.5$ for some positive values of $\delta_x$. Curves in the range 2 to 5 correspond to cases with $\delta_x=0.0; 0.2; 0.4$ and $0.45$, respectively. All these curves present a typical SMB that indicates that system suffers  only second-order phase transition (from the ferromagnetic phase to the paramagnetic phase). In particular, curve 2 ($\delta_x=0.0$) illustrates the influence of anisotropy $\delta_y$ on the magnetization behavior, making it lower than in the pure case at $\tau =0$.  Furthermore, it dramatically lowers the critical temperature compared to the pure case.  It is worth mentioning that small positive anisotropy values favor states with smaller spins, which explains these what explains these two these two occurrences. On the other hand, curves (3-5) demonstrate that the effect of $\delta_x$, at $\tau =0,$ leads to an increase in magnetization, bringing it closer to the value observed in the pure case. The anisotropy $\delta_x$ also makes the critical temperature decrease.
\par
On the other hand, in curves 6, 7  and 8, the magnetization appears in a region with behavior that differs from that of the SMB. Initially, the behavior follows the SMB in the low-temperature region, indicating a second-order phase transition. However, within a narrow temperature range {($\tau = 0.2106$ to $\tau = 0.5119$)}, the magnetization behavior diverges from that of the SMB, signaling a first-order phase transition within the ordered phase, that is, the system passes from an ordered phase with spins $S=3/2$ to another ordered phase with spins $S=1/2$, and then returns to the SMB, the magnetization goes to zero, marking a second-order phase transition between the ordered and disordered phases. Thus, we can consider that the competition between temperature and the two anisotropies in the spin system leads to a departure from the SMB, ultimately explaining the emergence of the first-order transition within the ordered phase. 
\par
Moreover, within Figure (1a), curves 9-11 exclusively display the characteristic behavior of SMB, signifying that solely second-order phase transitions take place in the system within this range of anisotropy values.
\\
\par
{\bf  Case $\delta_x$ negative:} 
Figure (1b) illustrates the magnetization behavior in relation to temperature, maintaining a constant value of $\delta_y=0.5$ for various negative values of $\delta_x$. Curves 1 to 7 represent scenarios with $\delta_x= 0.0, -0.2, -0.4, -0.6, -0.8, -1.0, -1.2$, respectively. All these curves exhibit the characteristic behavior of SMB, signaling that the system exclusively experiences a second-order phase transition between the ordered and disordered phases. Curve 8 represents the pure case (Spin-5/2 Ising model) and is shown as a guide for the eye. In particular, curve 1 ($\delta_x=0.0$), like curve 2 (in Figure 1), illustrates the influence of anisotropy $\delta_y$ in magnetization values, which makes it inferior to the pure case at $\tau =0$.
\par
At $\tau = 0.0$, curves 2 to 7 reveal a reduction in magnetization as $\delta_x$ increases, thereby distancing the magnetization from the ideal model (depicted by the dashed line). This observation signifies that negative values of $\delta_x$ promote the prevalence of states characterized by smaller spins. Consequently, as $\delta_x \rightarrow -\infty$, all spins converge to the smaller state of $S=1/2$. This fact implies that, in this limit, the system behaves like the Spin-1/2 model. 
\par
Figure (1b) also illustrates that increasing the magnitude of anisotropy, denoted as $\delta_x$, leads to the system attaining a higher critical temperature. As observed, curve 7 manifests the highest critical temperature, represented by ${\tau_c = 2.583}$. Due to this fact, all these curves (1-7) intersect at a specific temperature ($\tau_x$) which is lower than the critical temperature ($\tau_x < \tau_c$). 
\par 
This figure (1b) also  shows that the magnetization behavior for $\delta_y = 0.5$ and $\delta_x \leq 0$ is in accordance with the SMB type, suggesting that under these anisotropy, the system only undergoes phase transition ( ordered-disordered) exclusively of second order.  Curve 7 ($\delta_x = - 1.2$), at $\tau = 0$ presents the magnetization furthest from that of the pure case (smaller magnitude), and on the other hand it presents the highest critical temperature. Therefore, figure 1(b) shows that the greater the magnitude of $\delta_x$, the greater the magnetization departure from the pure case at $\tau = 0$. This fact explains the crossing of the other curves at a certain lower temperature $\tau_x$. that $\tau_c$. Finally,  the critical temperatures of the system under the effects of the two anisotropies  are lower than in the pure case ($\tau \leq 2.917$ ). 
\subsection{ Magnetization versus Temperature: Case $\delta_y=0.2$  Fixed }
{\bf  Case $\delta_x$ positive:} 
In the figure (1c), the magnitude of the anisotropy $\delta_y$ is reduced. This fact indicates that at $\tau=0$, the magnitude of the magnetization is closer to the pure case. 
This can be seen in curve 2 ($\delta_x=0.0$), it illustrates the influence of only $\delta_y$  anisotropy $\delta_y$ on the magnetization behavior, which the departure from the pure case is imperceptible at $\tau =0$.  On the other hand, as the temperature increases, the magnetization continually goes to zero, but reaches a critical temperature higher than in the previous case ($\delta_y =0.5$). This fact indicates that greater anisotropy is necessary to cause the system to undergo phase transition at lower critical temperatures.
Curves 3 and 4 present SMB-like behavior, indicating that the system only undergoes a second-order phase transition.  In particular, the curve 1 illustrates the case $\delta_x = \delta_y = 0$ representing the Pure Case - Spin-5/2 Ising Model, which is presented only as a guide for the eye. Curve 5 ($\delta_x =0.62$), exhibits SMB-like behavior at low temperatures, where the magnetization continually decreases from its maximum value ($m=2.5$), at $\tau =0$, until it reaches $ m=1.5$ in {$\tau= 0.4303$}. It then remains at this level, $ m=1.5$, until it reaches $\tau =0.5998 $ from where it drops abruptly to the value of $m=0.5$ at $\tau = 0.5$. This fact indicates a first-order phase transition between two ordered phases, that is, from the ordered phase with a predominance of $S=3/2$ spins to the ordered phase with $S=1/2$ spins.
\par
In summary, the behavior shown in figure (1c) is qualitatively the same as that shown in figure (1a). The behavior of magnetization in relation to temperature indicates that the system undergoes phase transition between second order phase (ordered - disordered phase), in addition to the first order transition between ordered phases.
\\
\par
{\bf  Case $\delta_x$ negative:} 
Figure (1d) illustrates the magnetization behavior in relation to temperature, maintaining a constant value of $\delta_y=0.2$ for  negative values of $\delta_x=  0.0, -0.2, -0.4, -0.6, -0.8, -1.0, -1.2$  corresponding to curves 1, 2, 4, 5, 6, 7, and 8 respectively. Here the smallest magnitude of $\delta_y$ brings the magnetization values closer to the value of the pure case at $\tau=0$, as shown in figure (1b).
All these curves ($1,2,4,5,6,7,8$) exhibit the characteristic behavior of SMB, signaling that the system exclusively experiences a second-order phase transition between the ordered and disordered phases. As a guide for the eye, the  curve 3 represents the pure case (Spin-5/2 Ising model). The curve 1 ($\delta_x=0.0$) only displays the influence of anisotropy $\delta_y$ on the magnetization behavior as a function of temperature, which indicates a departure in the magnitude of the magnetization from the value of the pure case at $\tau =0$. The other curves (2, 4, 5, 6, 7 and 8) show a greater distance, respectively. That is, the greater the magnitude of negative $\delta_x$, the greater this departure from the pure case at $\tau =0$. On the other hand, the system reaches higher critical temperatures as the magnitude of $\delta_x$ increases. Thus, for $\delta_x \geq 0.4$ the critical temperature of this system is greater than the critical temperature of the pure case ($\tau \geq 2.917$ ). Due to these facts, all these curves (1-8, except 3 ) intersect at a specific temperature ($\tau_x$) which is lower than the critical temperature ($\tau_x < \tau_c$), but greater than the $\tau_x$ of the case $\delta_y=0.5$. 
\subsection{ Magnetization versus Temperature: Case $\delta_y=0.0$   Fixed }
\begin{figure}[htb]
	\centering
	\includegraphics[scale=0.5, angle=-00]{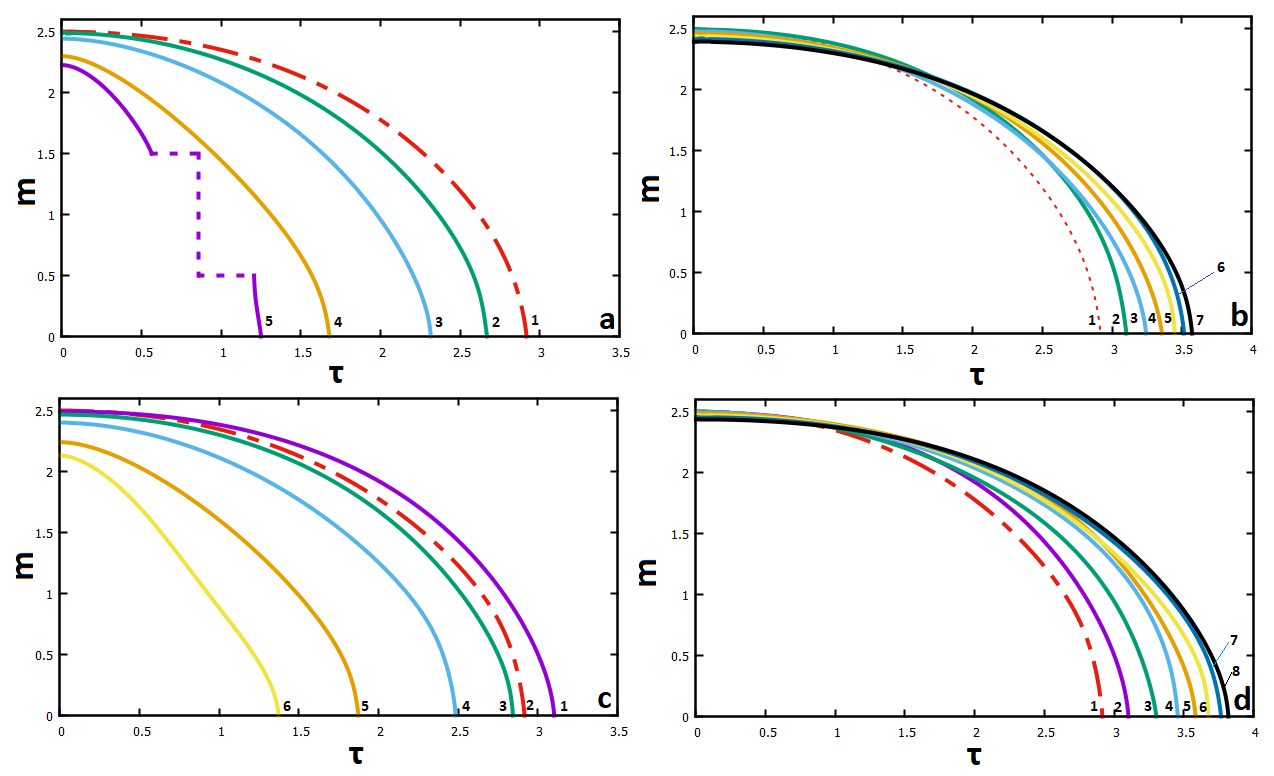}
	\caption{\it Diagram of magnetization versus temperature in ($m - \tau$) plane. Figure (a) presents the effects of single-ion anisotropies with $\delta_y = 0.0$ fixed and $\delta_x$ varying. Curve 1 ($\delta_y = 0.0, \delta_x = 0.0$), represented by a dashed–dotted line, corresponds to the pure model. The curves from 2 to 6 ($\delta_x = 0.2, 0.4, 0.6, 0.65$) represent the effects of anisotropy. Figure (b) - Curve 1 ($\delta_y = 0.0, \delta_x = 0.0$), represented by a dashed–dotted line, corresponds to the pure model. The curves from 2 to 7 ($\delta_x = -0.2, -0.4, -0.6, -0.80, -1.0, -1.2$) represent the effects of anisotropy.  Figure (c)  presents the effects of single-ion anisotropies with $\delta_y = -0.2$ fixed and $\delta_x$ varying. The pure case is represented by curve 2. The curves from 1 to 3-6 ($\delta_x = 0.0, 0.2, 0.4, 0.67$) represent the effects of anisotropy. Figure (d) presents the effects of single-ion anisotropies with $\delta_y = -0.2$ fixed and $\delta_x$ varying. The pure case is represented by curve 1. The curves from 2 to 8 ($\delta_x = 0.0, -0.2, -0.4, -0.6, -0.8, -1.0, -1.2$) represent the effects of anisotropy. Solid lines indicate standard magnetization behavior (SMB) and are related to a second-order phase transition. The dashed lines represent first-order phase transition within the ordered phase.}
\label{f1}
\end{figure}
{\bf  Case $\delta_x$ positive:} 
Figure (2a) displays the behavior of magnetization versus temperature, in plane ($\tau - m$) plane, with $\delta_y = 0.0$ fixed, and for various positive values of $\delta_x = 0.0,\,0.2 ,\, 0.4, \,0.6,\, 0.65$. Curve 1 shows the pure case $\delta_y = 0.0, \delta_x = 0.0$ (eye guide).
Curves 2 to 4, shown as solid lines ($\delta_x = 0.2,\,0.4,\, 0.6$), exhibit the standard magnetization behavior in the second-order phase transition region (SMB). At $\tau = 0$, the maximum value of magnetization ($m = 2.489; \,2.443;$ and $ 2.300$) shown by curves 2, 3 and 4, respectively. Increasing temperature induces the magnetization to continually go to zero at the critical temperature  ($\tau_c = 2,668;\, 2,316;\, 1,680$)  as shown by the three curves, exhibiting the SMB-like behavior.
Curve 5 ($\delta_x = 0.65$) shows the maximum value of magnetization ($m = 2.226$ ) at $\tau =0$, which drops continuously until reaching $m=1.5$ at $\tau =0.51$ , showing typical behavior of the second-order phase transition region. The magnetization remains at this value $m=1.5$ until it reaches $\tau =0.87$, where the magnetization suddenly drops to $m=0.5$. In this region, magnetization reveals a first-order phase transition within the ordered phase, indicating a phase transition from the ordered phase with spins $S=3/2$ to the ordered phase with spins $S=1/2$. The magnetization remains in this state until the temperature reaches $\tau = 1.74$ and then goes to zero continuously at the critical temperature $\tau = 1.254$, exhibiting SMB-like. 
\newline
Curves (1-5) indicate that the anisotropy $\delta_x$ causes the magnetization to decrease at $\tau= 0$. Here only the exchange interaction between the first neighboring spins $J > 0$ and anisotropy $\delta_x$ are present. On the one hand, the interaction $J > 0$ acts to leave the spins ordered in the ferromagnetic phase. On the other hand, $\delta_x$ acts to disorder the spins, causing the magnetization, at $\tau= 0$, to decrease with the increase in $\delta_x$.
\par
In summary in this regime, the system does not present tricritical behavior, that is, there is no first-order phase transition and tricritical points, the diagram only displays second-order phase transition lines.
\\
\par
{\bf  Case $\delta_x$ negative:} Figure (2b) shows the behavior of magnetization versus temperature in ($m- \tau$) with $\delta_y = 0$ plane and fixed for various negative values from $\delta_x = 0.0$ to $- 1.2$ with a step of $-0.2$. Curve 1 (dashed  line) displays the pure case (eye guide), on the other hand,  curves 2 to 7 (solid lines) exhibit the standard magnetization behavior (SMB), indicating that the system undergoes only a second-order phase transition. At $\tau =0$, curves 2 to 7, associated with the anisotropies $\delta_x = - 0.2$ to $\delta_x = - 1.2$, present maximum magnetizations close to the value of the pure case ($m=2.5$ ), that is, in terms $(m, \delta_x)$, we have $(2.4936, -0.2)$, $(2.4788, -0.4)$, $(2.4598, - 0.6)$, $(2.4383, - 0.8)$, $(2.4157, -1.0)$ and $(2.3924, -1.2)$ , respectively. In this case ($\tau =0$), only the exchange interaction (J > 0) and transverse single-ion anisotropy $\delta_x$ with negative values are acting on the system's spins, indicating that the anisotropy contributes to disordering the system.  When the temperature enters the competition ($J $ versus $\tau$ and $\delta_x$), the magnetization leaves its maximum value, decreasing continuously until it reaches the critical temperature $T_c$. Now in terms of $(\tau_c, \delta_x)$, curves 2 to 7 display $(3.1030, - 0.2)$, $(3.2444, - 0.4)$, $(3.3580, - 0.6)$, $ (2.4383, - 0.8)$, $(3.5159, - 1.0)$ and $(3.5744, - 1.2)$ , respectively.
\par
 Negative anisotropy favors states with spins whose magnitudes are smaller. So, at $\tau= 0$, all spins are ferromagnetically aligned due to the exchange interaction. Therefore, the action of transverse anisotropy explains the departure of the magnetization from the pure case, thus the greater departure for $\delta_x = - 1.2$, however small compared to the positive case. On the other hand, in the region of high temperatures (critical temperature region), on the contrary, $\delta_x = - 1.2$ contributes to the magnetic order, requiring greater thermal energy to disorder the magnetic system. This fact indicates that the lines representing greater and lesser transverse anisotropies $\delta_x$ intersect at a given temperature in the interval between $\tau_0 < \tau_x < \tau_c$. Thus, leading to critical temperatures much higher than the critical temperature of the pure case.

\subsection{ Magnetization versus Temperature: Case $\delta_y= - 0.2$   Fixed }
Figures (2c and d) show the behavior of magnetization versus temperature in  $(m - \tau\ $) plane for a fixed transverse anisotropy 
$\delta_y = - 0.2$ and several positive and negative values of the transverse anisotropy $\delta_x$, respectively. In both figures, the magnetization behavior shows that the spin system presents the standard behavior (SMB) of the region where the second-order phase transition occurs, as shown by curves 2 e 1 referring to the pure case (eye guide ), respectively.
\\
\par
{\bf  Case $\delta_x$ positive:}  Figure (2c) shows the cases in which $\delta_x$ assumes the positive values given by $\delta_x= 0.0,\, 0.2,\, 0.4,\, 0.6,\, 0.67$, corresponding to the curves $1, 3-6.$  Curve 1 indicates that the spins are only affected by transverse anisotropy $\delta_y$, as it corresponds to transverse anisotropy $\delta_x = 0$. At $\tau = 0$, only the exchange interaction $(J > )$ and $\delta_y$ compete to order and disorder the spin system. Therefore, as shown in the diagram, transverse anisotropy causes a small number of spins to lose ferromagnetic alignment, so the system acquires a maximum magnetization of $m= 2.4936$ with a small deviation from the pure case.  However, when thermal agitation enters the competition, the transverse anisotropy $\delta_y$ starts to order the system, as it causes the system of spins to become disordered at a much higher temperature than the pure case, that is, greater thermal energy is required so that the phase transition occurs at a critical temperature $\tau_c = 3.104$, greater than that of the pure case $\tau_c = 2.9167.$ Therefore, curve 1 intersects curve 2 at a temperature around {$\tau_x = 0.4491$.}
\newline

Curves (3-6) indicate the magnetization behavior with the transverse anisotropies $\delta_y$ and $\delta_x$ acting on the system's spins. These curves present the magnetization behavior (SMB), indicating that the system does not present tricritical behavior. At $\tau =0$, the magnetizations for each value of the anisotropy $(m, \delta_x)$ the figure shows are $(2.4673, 0.2)$, $2.4014, .0.4)$, $2.2416, 0.6)$ and $2.13274, 0.67)$, indicating a departure from the magnetization of the pure case ($m=2.5$). Here, the greater the value of $\delta_x$, the greater the deviation. On the other hand, when the temperature increases, the magnetization goes to zero at the critical temperature, and the curves indicate that $\delta_x > 0$ contributes significantly to disordering the system. Thus, the greater the magnitude of $\delta_x$ the lower the critical temperature, so curves 3 to 6 display $(\tau_c, \delta_x)$ given respectively by $(2.8442, 0.2)$, $(2.4839 , 0.4)$, $(1.8744, 0.6)$, $(1.3726, 0.67)$. These facts indicate that the anisotropy $\delta_x >0$ strongly contributes to disordering the system, that is, with its presence the need for thermal energy to disorder the system is smaller.
\\
\par
{\bf  Case $\delta_x$ negative:}
Figure (2d) shows the behavior of magnetization versus temperature in plane ($m- \tau$) plane with $\delta_y = -0.2$  fixed for various negative values from $\delta_x = 0.0$ to $- 1.2$ with a step of $-0.2$. Curve 1 (dashed  line) displays the pure case (eye guide), on the other hand,  curves 2 to 8 (solid lines) exhibit the standard magnetization behavior (SMB), indicating that the system undergoes only a second-order phase transition. This case is qualitatively the same as that presented in figure 2b with $\delta_y=0$. Here quantitatively, it shows an almost imperceptible deviation from the pure case in $\tau=0$, for example, in terms of $(m,  \delta_x)$, we have $(2.5, - 0.2)$, $ (2.4956, -0.4)$, $(2.4852, - 0.6)$, $(2.4383, - 0.8)$, $(2.4547, -1.0)$ and $(2.4369, - 1,2)$ , respectively.
On the other hand, the magnetization goes to zero at a much higher critical temperature, that is, in terms of $(\tau_c , \delta_x)$, curves 1 to 7 display $(3.3024, -0.2)$, $(3.4576, -0 ,4)$, $(3.5857, -0.6)$, $(3.6771,-0.8)$, $(3.7683, -1.0)$ and $(3.8206, -1.2)$, respectively. Therefore, we reach the same conclusion, $\delta_x <0$ contributes to ordering the system, it requires greater thermal energy to disorder the spin system. Furthermore, the system represented by the Blume-Capel Quantum Model of Spin $S = 5/2$ does not present tricritical behavior.
\subsection{Magnetization versus Temperature: Case $\delta_y= - 0.5$ Fixed}
   {\bf  Case $\delta_x$ positive:} 
  Figure (3a) illustrates the magnetization behavior in the $(m - \tau)$ plane for a fixed single-ion transverse anisotropy $\delta_y = -0.5$, with several positive values of $\delta_x$ ($0.0, 0.2, 0.4, 0.6, 0.67$).

Curve 1 ($\delta_x = 0$) represents the case where all spins are subject to a transverse anisotropy of magnitude $\delta_y = -0.5$. The pure case ($\delta_x = 0$, $\delta_y = 0$) is shown as a guide for the eye in curve 3. In this pure case, at $\tau = 0$, the exchange interaction ($J > 0$) ferromagnetically aligns all spins in the $S = 5/2$ state, yielding the maximum magnetization $m = 2.5$. Under the influence of transverse anisotropy ($\delta_y$), the maximum magnetization is reduced compared to the pure case. For curve 1, the magnetization drops to $m = 2.4697$. As temperature increases, the magnetization decreases, crossing curve 3 (the pure case), and eventually reaches the critical temperature $\tau_c = 3.3080$, which is higher than that of the pure case. At $\tau_c$, the magnetization vanishes ($m = 0$), signaling a second-order phase transition from the ferromagnetic to the paramagnetic phase.

Curve 2 ($\delta_x = 0.2$) exhibits a similar trend: the maximum magnetization is $m = 2.4259$, again deviating from the pure case, but the critical temperature ($\tau_c = 3.0312$) is lower than that of curve 1. This indicates that $\delta_x$ induces disorder, while $\delta_y < 0$ contributes to maintaining order. Curve 2 also crosses curve 3, which corresponds to the pure spin-$5/2$ Blume-Capel quantum model.

The remaining curves (4–6) display the same features as those shown in Figure (2c). These curves show larger deviations in magnetization from the pure case and reach critical temperatures lower than that of the pure spin-$5/2$ Blume-Capel quantum model (curve 3).

In summary, these results demonstrate that $\delta_y < 0$ helps preserve order in spin systems, raising the critical temperature. Conversely, $\delta_x > 0$ strongly promotes disorder, leading to greater magnetization deviations from the pure case at $\tau = 0$ and yielding lower critical temperatures than the pure case when $\delta_x > 0.2$.
  \par
   {\bf  Case $\delta_x$ negative:} The figure (3b) shows the behavior of magnetization versus temperature in plane ($m- \tau$) plane with $\delta_y = -0.5$  fixed for various negative values from $\delta_x = 0.0$ to $- 1.2$ with a step of $-0.2$. Basically, this diagram presents the same magnetization behavior in relation to temperature shown in figure (2d). There are only some quantitative differences in the magnetization values at $\tau=0$ ), and in the critical temperatures. For example, now the magnetizations $\tau=0$ ($m_{\tau_0}$), in terms of $(m_{\tau_0}, \tau_c, \delta_x)$, we have$(2.4697, 3.3079, 0.0)$, $(2.4910, 3.5238, - 0.2)$, $ (2.4991, 3.7037,  -0.4)$, $(2.4993, 3.8312, - 0.6)$, $(2.4942, 3.962, - 0.8)$, $(2.4859, 4.0435, -1.0)$ and $(2.4752, 4.1137, - 1,2)$ represented by curves 2-8, respectively.  
On the other hand, all curves exhibit standard magnetization behavior (SMB), indicating that the system undergoes only a second-order phase transition.   
 The figure leads us to conclude that $\delta_x <0$ contributes to ordering the system, requiring the system to spend more thermal energy to go to the disordered phase.   Therefore, this spin system represented by the Blume-Capel Quantum Model of Spin $S = 5/2$ does not present tricritical behavior.  
\begin{figure}[htb]
	\centering
	\includegraphics[scale=0.5, angle=-00]{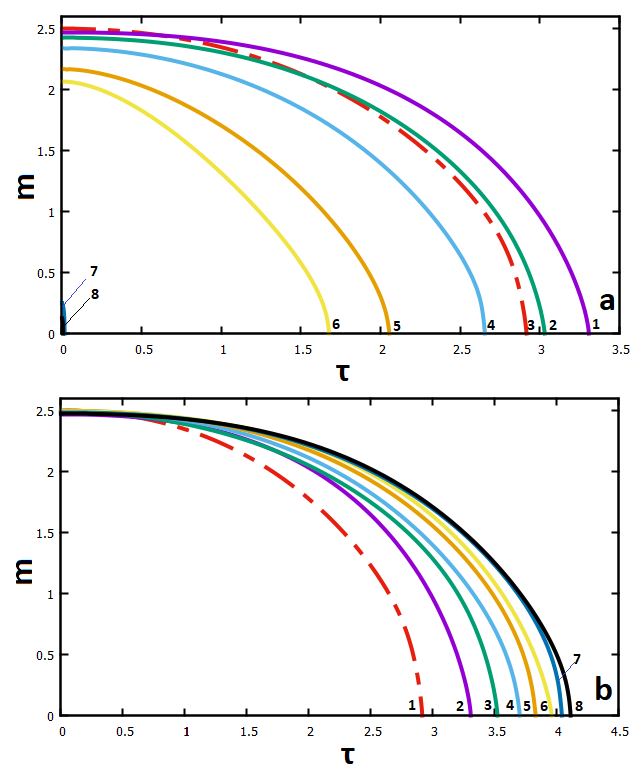}
	\caption{\it Diagram of magnetization versus temperature in ($m - \tau$) plane. Figure (a) presents the effects of single-ion anisotropies with $\delta_y = 0.5$ fixed and $\delta_x$ varying. Curve 1 ($\delta_y = 0.5 , \delta_x=0.0$) - dashed and dotted  line - pure model. The curves from 2 to 11 $\delta_x=0.0$ the anisotropies have $\delta_y = 0.5$ and $\delta_x = 0.0, 0.2, 0.4, 0.45, 0.5, 0.6, 0.7, 0.8, 1.0, 1.2$, respectively.  Figure (b) corresponds to $\delta y=0.5$ and negative $\delta$, varying $\delta_x = -1.2, -1.0, -0.8, 0.6, - 0.4, - 0.2, 0.0$ referring to curves 1 to 7, respectively. Figure (c) corresponds to $\delta_y=0.2$ and $\delta_x =
0.0, 0.2, 0.4, 0.6, 0.62$ linked to curves 1 to 5, respectively. Figure (d) corresponds to $\delta_y=0.2$ and $\delta_x =
0.0, -0.2, -0.4, -0.6, -0.8, -1.0, -1.2$ linked to curves 1, 2, 4, 5, 6, 7 and 8 respectively.
	 Solid lines indicate standard magnetization behavior (SMB) and are related to a second-order phase transition. The dashed lines represent first-order phase transition within the ordered phase.}
	\label{f1}
\end{figure}
\section{Conclusion}
In this work, the magnetization behavior in relation to temperature of a spin system, represented by the Blume-Capel model of $S=5/2$ with transverse anisotropy of a single ion, was studied  using mean field theory based on Bogoliubov's inequality for Gibbs free energy. This approximation provides the Gibbs free energy (fundamental equation) and magnetization (equation of state). 
\par
Competitions (order-disorder) in the spin system are associated with four ingredients present in the physical system. The first \'e $J > 0$, the exchange interaction which ferromagnetically connects all pairs of neighboring first spins, inducing a ferromagnetic order in the lattice. The second is temperature, which produces vibrations in the lattice, inducing disorder in the system, that is, it leads to the order-disorder phase transition. The third and fourth are associated with single-ion transverse anisotropies ($\delta_y$ and $\delta_x$), which favor states with spins $S = 5/2$ and $3/2$, when the value of parameter D is positive ($ D>0$) or to the state with spin $S=1/2$ when it has negative magnitude ($D <0$). 
\par
All diagrams showed that the Blume-Capel model of $S=5/2$ with transverse anisotropy of a single ion does not exhibit tricritical behavior, in any combination. Only for $\delta_y > 0$ does the model present a deviation from the standard behavior (SMB), exhibiting a first-order phase transition, within the ordered phase, as shown in figures 1(a, c) and figure (2a).
\par
Our contributions to the studies of this system consisted of showing that the system does not present tricritical behavior in all ranges of values of $\delta_y$ and $\delta_x$. This fact, and the other contributions, are related to the information contained in the very rich diagrams of magnetization behavior versus temperatures. 
\par
In summary, we can state that large values $\delta_y > 0$ and $\delta_x > 0$ cause the system to become disordered at temperatures lower than the pure case. On the other hand, decreasing the magnitude of $\delta_y > 0$ and with $\delta_x < 0$ increases the critical temperature, this requiring more thermal energy to clear the system. 
In the future, the treatment could occur using an improved variational approach or through quantum Monte Carlo simulation and other approximations, in addition to considering larger spins, which can be compared with experimental results.

\begin{acknowledgments}

The authors acknowledge the financial support from the Brazilian agencies UFMT, IFMT and CAPES.

\end{acknowledgments}


\begin{thebibliography}{90}
\bibitem{blu} M. Blume, Phys. Rev. {\bf 141} (1966) 517.

\bibitem{cap} H. W. Capel, Physica {\bf 32} (1966) 966.

\bibitem{exp0} W.B. Yelon, R. Scherm, Solid State Communications,  {\bf 15}, (1974) 39 .

\bibitem{exp1}   K. Strnat, IEEE Trans. Magn. {\bf 6}, 182 (1970).

\bibitem{exp2} J. Herbst, Rev. Mod. Phys. {\bf 63}, 819 (1991).

\bibitem{exp3} W. K\"orner, G. Krugel, and C. Els\"asser, Sci. Rep. {\bf 6} , 24686 (2016).

\bibitem{exp4} S. Suzuki, T. Kuno, K. Urushibata, K. Kobayashi, N. Sakuma, K. Washio, M. Yano, A. Kato, and A. Manabe, J. Magn. Magn. Mater. {\bf 401}, 259 (2016).

\bibitem{exp5} Y. Harashima, K. Terakura, H. Kino, S. Ishibashi, and T. Miyake, Phys. Rev. B {\bf 92}, 184426 (2015).

\bibitem{exp6} Y. Hirayama, Y. Takahashi, S. Hirosawa, and K. Hono, Scr. Mater. {\bf 95}, 70 (2015).

\bibitem{exp7} T. Miyake, K. Terakura, Y. Harashima, H. Kino, and S. Ishibashi, J. Phys. Soc. Jpn. {\bf 83}, 043702 (2014).

\bibitem{exp8} J. Coey, IEEE Trans. Magn. 47, 4671 (2011).

\bibitem{exp9}  J. M. D. Coey, Rare-Earth Iron Permanent Magnets (Oxford University Press, Oxford, 1996), Vol. {\bf 54}.

\bibitem{clau1} Salgado, CM ; de Carvalho, NL; de Arruda, PHZ; Godoy, M; de Arruda, AS; Costabile, E; de Sousa, JR; Physica A {\bf 522} ( 2019) 18. 

\bibitem{seto1}  G. Seto, A. Kpadonou, R. Yessoufou, E. Albayrak, Physica B: Condensed Matter
 {\bf 634}, 1  {2022}, 413782. 

\bibitem{ita0} Itacy J. Souza, M. Godoy, A. S. de Arruda and T.M. Tunes, Eur. Phys. J. B  {\bf 93} (2020) 215.

\bibitem{seto2} G. Seto, R.A.A. Yessoufou, A. Kpadonou, E. Albayrak,Physica A {\bf 604} (2022) 127939. 

\bibitem{181} P. V. Santos, F. A. da Costa, and J. M. de Ara\'ujo, J. Magn. Magn. Mater. {\bf 451} (2018) 737.    

\bibitem{171} G. Gulpinar, R. Erdem, and M. Agartioglu, J. Magn. Magn. Mater {\bf 439} (2017) 44.

\bibitem{172} M. Acharyya and A. Halder, J. Magn. Magn. Mater {\bf 426} (2017) 53.

\bibitem{173} W. P. da Silva, P. H. Z. de Arruda, T. M. Tunes, M. Godoy and A.S. de Arruda, J. Magn.
Magn. Mater. {\bf 422} (2017) 367.

\bibitem{174} S. Sumedha, N. K. Jana, J. Phys. A: Math. Theor. {\bf 50} (2017) 015003.

\bibitem{a2} A. S. de Arruda and  W. Figueiredo, Modern Physics Letters B {\bf 11} (21-22), (1997) 973.

\bibitem{cdd} C. De Dominicis, I. Giardina, Random Fields and Spin Glasses (Cambridge University Press, Cambridge, 2006).

\bibitem{a1} A. Weizenmann, M. Godoy, A. S. de Arruda, D. F. de Albuquerque, and N. O. Moreno,  Physica B {\bf 398} (2007) 297  2 (2007) 297. 
\bibitem{ku} M. Kaufman, P. E. Klunzinger, and A. Khurana, Phys. Rev. B {\bf 34} (1986) 4766.

\bibitem{k1} M. Kaufman and M. Kanner, Phys. Rev. B {\bf 42} (1990) 2378.

\bibitem{a3} J. B. Santos, N. O. Moreno, D. F. de Albuquerque, and A. S. de Arruda, Physica B {\bf 398} (2007) 294.

\bibitem{kenz} L. Bahmad, A. Benyoussef, and A. El Kenz, M. J. Condensed Matter {\bf 9} (2007) 142.

\bibitem{ita1} I. J. Souza, P. H. Z. de Arruda, M. Godoy, L. Craco, and A. S. de Arruda,  Physica A {\bf 444} (2016) 589.

\bibitem{ze1} J. S. da Cruz Filho, T. M. Tunes, M. Godoy, and A. S. de Arruda, Physica A {\bf 450} (2016) 180.

\bibitem{ze2} J. S. da Cruz Filho, M. Godoy, A. S. de Arruda, Physica A {\bf 392} ( 2013) 6247. 

\bibitem{mel} D. Mukamel and M. Blume, Phys. Rev. A {\bf 10} (1974) 610.

\bibitem{gul} Yenal Karaaslan, G{\"u}l G{\"u}lpinar, Journal of Magnetism and Magnetic Materials 652 (2026) 174185 .

\bibitem{ita3} Souza, I.J., de Faria, J.L.B, Bento, R.R.F., de Arruda, A.S., Tunes, T.M., de Arruda M.F.Z., Karimou, M., Phase Transitions {\bf 98} (2025) 505. 

\bibitem{so1} E. Costabile, J. R. Viana, J. R. de Sousa and A. S. de Arruda, Solid State Comm. {\bf 212} (2015) 30.

\bibitem{pla3} D. C. Carvalho and J. A. Plascak, Physica A {\bf 432} (2015) 240.

\bibitem{san1} P. V. Santos, F. A. da Costa and J. M. de Ara\'ujo, Physics Letters A  {\bf 379} (2015) 1397.

\bibitem{so2} J. R. de Sousa, N. Branco, Phys. Rev. E {\bf 77} (2008) 012104. 

\bibitem{falk} Falk H., Amer. J. Phys., 38 (1970), p. 858.


\end{thebibliography}
\end{document}